\documentclass[conference,letterpaper]{IEEEtran}
\IEEEoverridecommandlockouts

\usepackage{cite}
\usepackage{amsmath,amssymb,amsfonts}
\usepackage{graphicx}
\usepackage{textcomp}
\usepackage{xcolor}
\usepackage{enumitem}
\usepackage{multirow}
\usepackage{tabularx}
\usepackage{array}
\usepackage{algorithm}
\usepackage{algpseudocode}
\usepackage{algorithmicx}
\usepackage{makecell}
\usepackage{hyperref}
\algrenewcommand\algorithmicindent{0.8em}

\begin{document}


\title{Automated Extraction of Protocol State Machines from 3GPP Specifications with Domain-Informed Prompts and LLM Ensembles}

\author{
    Miao Zhang$^{1}$,
    Runhan Feng$^{2}$,
    Hongbo Tang$^{1}$,
    Yu Zhao$^{1}$,
    Jie Yang$^{1}$,
    Hang Qiu$^{1}$,
    Qi Liu$^{2}$
    \\
    \IEEEauthorblockA{
        $^{1}$Information Engineering University, China \\
        $^{2}$Purple Mountain Laboratories, China \\
        Email: 1152461073@qq.com, fengrunhan@pmlabs.com.cn,
        tahobo@sina.com, itsyz@foxmail.com \\
        yj\_csu@126.com, qiuhang\_ndsc@163.com, liuqi@pmlabs.com.cn
    }
}



\maketitle

\def\toolname{SpecGPT}
\newcommand\todo[1]{\textcolor{red}{\textbf{(ToDo: #1)}}}

\begin{abstract}
Mobile telecommunication networks are foundational to global infrastructure and increasingly support critical sectors such as manufacturing, transportation, and healthcare. The security and reliability of these networks are essential, yet depend heavily on accurate modeling of underlying protocols through state machines. While most prior work constructs such models manually from 3GPP specifications, this process is labor-intensive, error-prone, and difficult to maintain due to the complexity and frequent updates of the specifications. Recent efforts using natural language processing have shown promise, but remain limited in handling the scale and intricacy of cellular protocols. In this work, we propose SpecGPT, a novel framework that leverages large language models (LLMs) to automatically extract protocol state machines from 3GPP documents. SpecGPT segments technical specifications into meaningful paragraphs, applies domain-informed prompting with chain-of-thought reasoning, and employs ensemble methods to enhance output reliability. We evaluate SpecGPT on three representative 5G protocols (NAS, NGAP, and PFCP) using manually annotated ground truth, and show that it outperforms existing approaches, demonstrating the effectiveness of LLMs for protocol modeling at scale.
\end{abstract}

\begin{IEEEkeywords}
Large Language Models, State Machine, 3GPP Standards
\end{IEEEkeywords}

\section{Introduction}
Mobile telecommunication networks constitute a fundamental pillar of global communication infrastructure. According to the GSMA Mobile Economy Report \cite{gsma}, the number of unique mobile subscribers worldwide surpassed 5.8 billion in 2024. Beyond personal communication, mobile networks are increasingly integrated into critical sectors such as industrial manufacturing, transportation, and healthcare, rendering them an indispensable component of modern society. As such, their failure can lead to far-reaching and severe consequences. Ensuring the reliability and security of telecommunication networks have therefore become more crucial than ever \cite{khan2019survey}. 

Network protocols specify the rules for information exchange between entities in a network and serve as the foundation of telecommunication systems. To improve the overall security of telecommunication networks, researchers have proposed various approaches to evaluate protocol security from both design and implementation perspectives. For instance, formal methods are employed to verify the security properties of protocol designs \cite{akon2025control, basin2018formal, hussain20195greasoner, hussain2018lteinspector,cremers2019component}, while software testing techniques such as fuzz testing are applied to uncover vulnerabilities in protocol implementations \cite{karim2021prochecker, dongcorecrisis, yang2024feedback, bennett2024ransacked, kim2019touching,huang2024large}. The vast majority of these approaches, either directly or indirectly, rely on protocol state machines, which serve as a common modeling framework for representing and analyzing protocol workflows. 

However, obtaining state machine models for telecommunication network protocols is a non-trivial task. Specifically, these protocols are designed and maintained by the 3rd Generation Partnership Project (3GPP), a global consortium comprising network operators, equipment vendors, and researchers \cite{3gpp}. The standardized protocol details are documented in technical specifications written in semi-structured natural language intended for domain experts. Due to the involvement of a wide range of cellular stakeholders and network entities, the diversity of use cases, and strict backward compatibility requirements, 3GPP protocol standards are typically complex and verbose. They are distributed across multiple specification documents, each focusing on specific layers, procedures, or architectural components of the protocol, and often span hundreds or even thousands of pages. Since 3GPP does not provide formal models of the protocols, existing work typically relies on domain experts to manually interpret the specifications and construct corresponding state machines. This process is labor-intensive, time-consuming, and prone to human error. The accuracy and level of abstraction of such hand-crafted models significantly influence the effectiveness of downstream tasks such as protocol verification. Furthermore, 3GPP specifications undergo frequent updates—typically five to six times per year. Keeping pace with these changes is essential for assessing the security of the most recent protocol versions. However, due to the high time and labor costs involved, regularly updating manually constructed models is not feasible. In this work, we aim to automatically extract protocol state machines from 3GPP specifications.

In recent years, researchers have explored the use of natural language processing (NLP) techniques for the automated analysis of protocol documents \cite{chen2021bookworm, chen2022forest,nguyen2024cama}. Pacheco et al. \cite{pacheco2022automated} proposed RFCNLP which is capable of extracting finite state machines (FSMs) from Request for Comments (RFC) documents. However, compared to the relatively small-scale protocols defined in RFCs, cellular network protocols are significantly more complex, involving a larger number of state transitions, detailed protocol field descriptions, and intricate cross-layer interactions, which render RFCNLP ineffective. To address these challenges, Ishtiaq et al. \cite{al2024hermes} introduced Hermes, which leverages a deep-learning-based neural parsing model to perform constituency parsing of 3GPP documents and employs a domain-specific language to generate the final FSMs. Nevertheless, while Hermes incorporates domain-specific design elements, its effectiveness is still constrained by the inherent complexity of the task.

Since the release of ChatGPT in late 2022 \cite{chatgpt}, large language models (LLMs) have demonstrated remarkable natural language comprehension capabilities, leading to groundbreaking progress in a wide range of NLP tasks, including document summarization and structured data extraction \cite{zheng2025large,duclos2024utilizing,ma2024specgen,maklad2025retrieval,sharma2023prosper}. In this work, we propose leveraging large language models (LLMs) to automatically extract protocol state machines from 3GPP specifications. However, the application of LLMs to highly technical, domain-specific protocol specifications remains relatively unexplored. Applying LLM to infer protocol FSM from 3GPP specifications encounters several challenges. On the one hand, as is mentioned before, the 3GPP specifications as well as the protocols are highly complex. On the other hand, despite their strong language comprehension, LLMs are prone to hallucinations, especially when dealing with complex or ambiguous input \cite{huang2025survey, xu2024hallucination,xie2025effective}. 

To overcome these challenges and unlock the full potential of large language models, we introduce \toolname{}, a novel framework designed to automatically extract protocol FSMs from complex technical documents. The framework first segments 3GPP specifications into semantically meaningful paragraphs. It then constructs domain-informed prompts to guide the model toward accurate interpretations. Finally, ensemble methods are applied across multiple LLMs to aggregate and refine the results, enhancing both reliability and performance. We conduct a comprehensive evaluation of \toolname{} on 3GPP specifications, targeting three representative protocols: 5G NAS (Non-Access Stratum) \cite{5gnas}, NGAP (Next Generation Application Protocol) \cite{5gngap}, and PFCP (Packet Forwarding Control Protocol) \cite{5gpfcp}. Due to the absence of publicly available ground truth for their complete state machines, we construct manually annotated references. Experimental results indicate that \toolname{} achieves an F1-score of up to 91.14\% in the task of state transition extraction, outperforming the state-of-the-art method.

In summary, we make the following major contributions in this work:
\begin{enumerate}[nolistsep, labelindent=\parindent, leftmargin=*, label=$\bullet$]
\item We designed a novel technique that can automatically extract protocol state machines from 3GPP specification documents. Our approach leverages large language models to understand the natural language in the specifications, incorporates domain knowledge to construct prompts, and employs model ensemble techniques to further improve the performance of state machine construction.

\item  We implement our technique in a tool called \toolname{} and applied it to real-world 3GPP protocol specification documents for comprehensive evaluation. Experimental results show that our approach can accurately construct protocol state machine models and outperforms existing state-of-the-art solutions.
\end{enumerate}

The rest of the paper is organized as follows. We first introduce the background and research scope in Section \ref{background}. We then describe the design of \toolname{} in Section \ref{design}. The evaluation results are presented in Section \ref{evaluation}, followed by a discussion in Section \ref{discussion}. We introduce related work in Section \ref{related} and conclude our paper in Section \ref{conclusion}.
\section{Background}
\label{background}
\subsection{Cellular network}
Cellular networks have undergone a continuous evolution from 1G to 5G, transitioning from voice-centric systems to advanced, data-driven architectures capable of supporting high data rates, ultra-low latency, and massive machine-type connectivity. A canonical cellular network architecture comprises two principal components: the Radio Access Network (RAN) and the Core Network (CN). Driven by increasing demands for service agility, flexibility, and scalability, network functions have been progressively decoupled from dedicated hardware, fostering the adoption of modular, service-based architectures. 

\begin{figure}[htbp] 
    \centering
    \includegraphics[width=\linewidth]{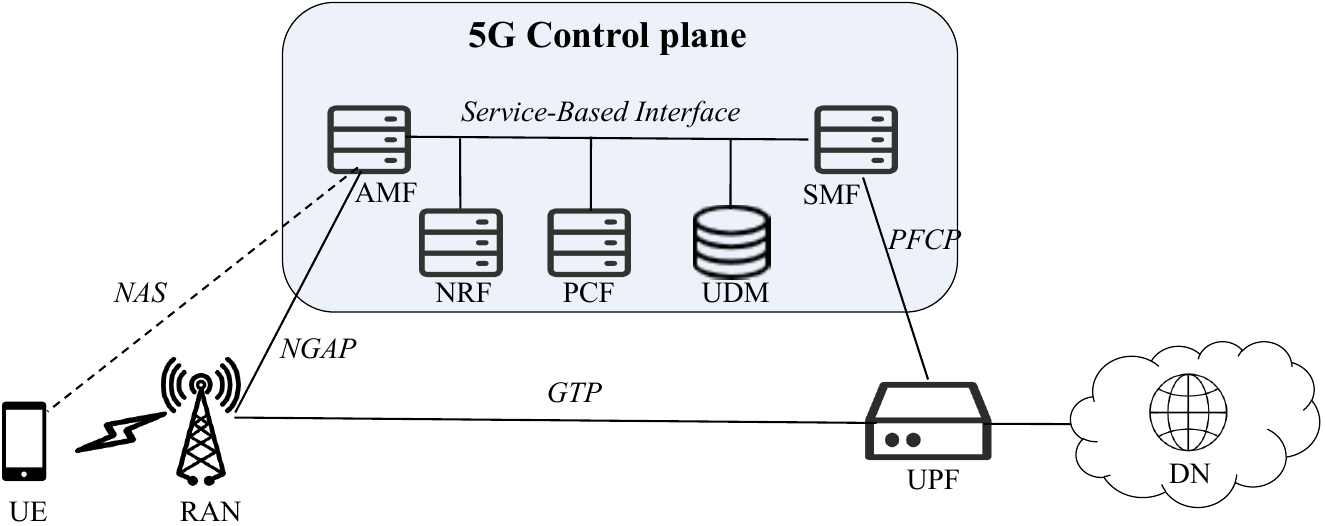} 
    \caption{5G Architecture.}
    \label{architecture diagram}
\end{figure}

As illustrated in Figure \ref{architecture diagram}, the 5G CN is logically decomposed into a set of independent network functions (NFs), each designed to fulfill a specific operational role. These NFs communicate with one another through standardized interfaces, each associated with a dedicated protocol stack that supports various control- and user-plane operations. For example, the NGAP facilitates control-plane signaling between the RAN and the Access and Mobility Management Function (AMF); the NAS protocol enables mobility management and session control between the User Equipment (UE) and the AMF; and the PFCP allows the Session Management Function (SMF) to govern user-plane behavior and enforce traffic handling policies on the User Plane Function (UPF).

\subsection{3GPP Specifications}
3GPP is a collaborative standards organization comprising multiple regional telecommunications standards development organizations, dedicated to the development of globally applicable technical specifications and technical reports for mobile communication systems. By defining standardized interface protocols, 3GPP ensures interoperability among equipment from different vendors, significantly promoting the development and standardization of the global mobile communications industry.

3GPP has developed a comprehensive suite of technical specifications that span multiple layers of the cellular network architecture, including the RAN, CN, and UE. The volume of released specifications is substantial, and the standards are continuously refined to accommodate emerging technologies and evolving use cases. For instance, in the context of 5G, since the introduction of 5G New Radio in Release 15, the specifications have evolved through successive releases up to Release 19. Each new release introduces enhancements over its predecessor and extends support for novel application scenarios, such as integrated sensing and communication.

Notably, each technical specification often spans hundreds of pages, addressing a wide range of deployment scenarios and operational conditions in mobile networks. This extensive and ever-growing body of documentation makes it increasingly inefficient and time-consuming for engineers and researchers to manually understand and interpret the standards. Therefore, there is a pressing need to explore automated approaches for the efficient parsing and application of these technical specifications.

\section{Design}
\label{design}
Automatically extracting protocol state machines from 3GPP specifications is a highly non-trivial task. In this section, we present the design of \toolname{}, our proposed framework for this challenge. We employ large language models (LLMs) to assist in interpreting the specifications written in natural language. Furthermore, we incorporate domain knowledge into prompt design and adopt an ensemble of LLMs to enhance overall performance and robustness.

\subsection{FSM Statement}
The specification of a cellular network state machine is typically defined as a quintuple \( \langle Q, \Sigma, q_{0}, \delta, F \rangle \). The aim of \toolname{} is to construct a formal model as follows:

\begin{itemize}
    \item \textbf{State Set (\( Q \))}: A finite set of states representing all possible conditions that may occur within the protocol.
    \item \textbf{Input Alphabet (\( \Sigma \))}: Also referred to as the set of input symbols, this includes all possible transitions in the protocol implementation. Each transition represents a specific condition or action within the protocol, thereby triggering state transitions.
    \item \textbf{Initial State (\( q_{0} \))}: The set of initial states, which is a subset of the state set \( Q \) (\( q_{0} \in Q \)), indicating the state in which the protocol resides at startup.
    \item \textbf{State Transition Function (\( \delta \))}: A rule that defines how to transition to the next state based on the current state and input symbol. Formally, this is understood as \( \delta: Q \times \Sigma \rightarrow Q \).
    \item \textbf{Final State (\( q_{f} \))}: The set of final states, which is a subset of \( Q \) (\( q_{f} \in Q \)), indicating the terminal states when the protocol has completed its transitions.
\end{itemize}

\subsection{Overview}
\begin{figure*}[htbp] 
    \centering
    \includegraphics[width=1.0\textwidth]{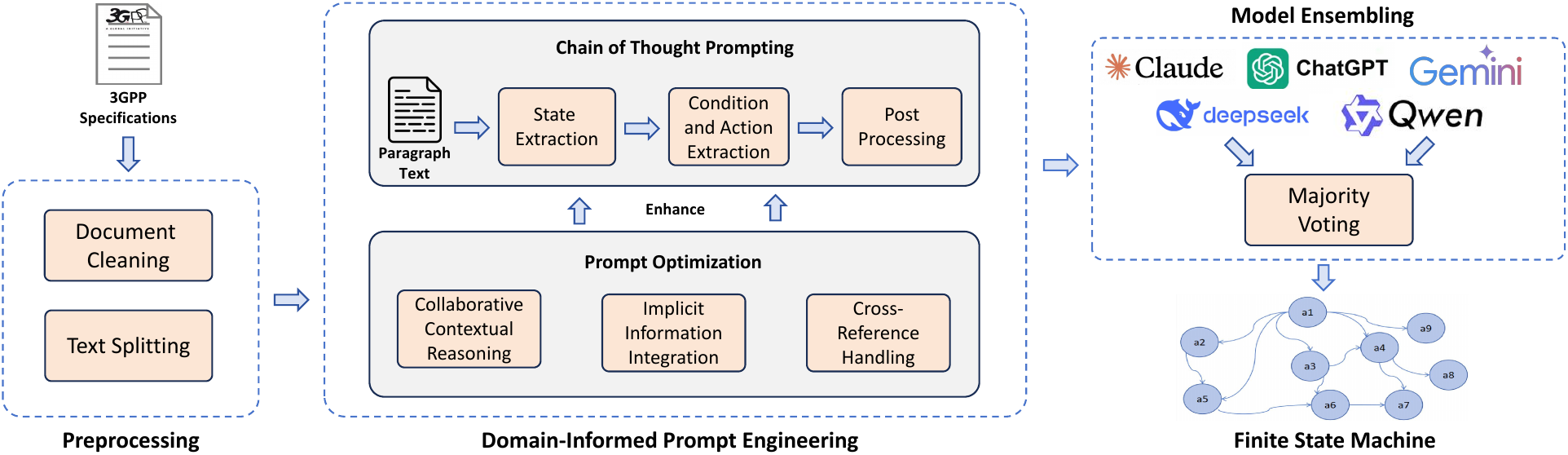}
    \caption{Overview of SpecGPT}
    \label{Overview of SpecGPT}
\end{figure*}

We propose a framework named \toolname{} for automatically extracting state machines from 3GPP protocol specifications using LLMs. As illustrated in Figure \ref{Overview of SpecGPT}, the framework consists of three main components. In the \textbf{preprocessing stage}, we first perform text cleaning on the original 3GPP documents. We then merge related sections to restructure the content, enabling the documents to be segmented into smaller, coherent chunks suitable as inputs for the LLMs. This mitigates issues such as input truncation due to sequence length limitations. The preprocessing procedure is described in detail in Section \ref{Preprocess}. In the \textbf{prompt engineering stage}, we design Chain-of-Thought \cite{wei2022chain} prompts to guide the automatic extraction of protocol state machines. Additionally, we incorporate domain-specific knowledge to comprehensively refine and optimize the prompts, thereby improving their effectiveness. Further details are presented in Section \ref{Prompt Engineering}. In the \textbf{model ensembling stage}, we utilize multiple LLMs in parallel to analyze the protocol specifications. Their outputs are then aggregated using a majority voting strategy to construct the final finite state machine. The ensembling process is detailed in Section \ref{Ensembling}.

\begin{algorithm}[htbp]
\footnotesize
\caption{Section-level Paragraph Window Merging}
\label{merging algorithm}
\begin{algorithmic}[1]
\Require Protocol document $D$
\State $sections \gets \text{extract\_section\_numbers}(D)$
\State $paragraph\_to\_section \gets \text{map\_paragraphs\_to\_sections}(D, sections)$
\State $section\_tree \gets \text{build\_section\_tree}(sections)$
\For{node $\in\ \text{get\_leaf\_nodes}(section\_tree)$}
    \State $parent \gets \text{get\_parent}(node, section\_tree)$
    \If{$\text{all\_children\_merged}(parent) == \text{False}$}
        \State $content \gets \text{merge\_content}(\text{get\_children}(parent))$
        \State $\text{assign\_content}(parent, content)$
        \State $\text{set\_all\_children\_merged}(parent) \gets \text{True}$
    \EndIf
\EndFor
\State $merged\_windows \gets \text{get\_parent\_nodes\_with\_content}(section\_tree)$
\Return $merged\_windows$
\end{algorithmic}
\end{algorithm}

\subsection{Preprocess}
\label{Preprocess}
The input to the preprocessing stage is the raw 3GPP specification. After text cleaning and section consolidation, the output consists of small segments that remain aligned with the original document structure.

\paragraph{\textbf{Document cleaning}} The 3GPP specifications are provided in Word format and contain a significant amount of non-body content, such as tables of contents, headers, footers, superscripted footnote markers, redundant blank lines, and fragmented or irrelevant chart data. Such noisy text can hinder the model’s comprehension and reduce reasoning accuracy. To address this, we apply regular expressions to automatically detect and remove residual elements, including tables of contents and headers.

\paragraph{\textbf{Text Splitting}} Due to the context length limitations of large language models, it is necessary to segment the document content before processing. Traditional strategies typically adopt either fixed-size sliding windows or paragraph-based splitting. However, the former may lead to context loss due to truncation, while the latter often results in overly fine-grained segments, causing a large number of ineffective queries (e.g., some paragraphs in 3GPP documents contain only one or two sentences). To address this, we perform section consolidation leveraging the hierarchical section structure of 3GPP specifications to produce semantically coherent input chunks. Specifically, we define a hierarchical relationship between sections as parent–child (also referred to as parent and leaf nodes in our algorithm). As shown in Algorithm \ref{merging algorithm}, the proposed approach first parses section numbers and formatted markers to accurately extract and map section nodes to their corresponding paragraphs. Furthermore, a section tree is constructed based on section numbering, clarifying the hierarchical and parent–child relationships across various section levels. On this basis, for the lowest-level leaf nodes of the section tree, a bottom-up merging strategy is applied: all leaf node contents under the same parent are merged recursively into their parent node, thereby achieving hierarchical content aggregation across the document.

\subsection{Prompt Engineering}
\label{Prompt Engineering}
In this stage, we integrate domain knowledge with chain-of-thought prompt engineering to decompose the FSM extraction task into manageable sub-tasks. We further design targeted prompt optimization strategies to maximize the potential of large language models.

\paragraph{\textbf{Chain-of-Thought Prompting}}
The finite state machine extraction task is divided into three main stages: state extraction, transition extraction, and post-processing. In our formulation, each transition is characterized by an associated action and condition. The details of each step are presented in the following subsections.


\textbf{State Extraction.} As shown in Figure \ref{architecture diagram}, the 3GPP specifications define a wide range of protocols, each designed for different usage scenarios. As a result, they vary in both design and descriptive style in the specifications. Based on our in-depth analysis of the 3GPP protocol specifications, we categorize them into two types: state-oriented and procedure-oriented protocols.

State-oriented protocols are those in which transition nodes have clearly defined states, while procedure-oriented protocols feature transition nodes described in terms of procedural steps. Based on this classification, we divided prompts into state prompts and process prompts during state extraction. For state-oriented protocols (e.g., NAS), state names are explicit— for example, ``5GMM-REGISTERED" indicates that the UE has successfully registered with the core network, representing a clear state. Here, the prompt can be formulated as ``extract the states mentioned in the text." In contrast, for procedure-oriented protocols (e.g., PFCP), states typically refer to nodes within a procedure, such as ``Established connection with SMF." In such cases, we focus on whether a specific procedure has been completed, and treat this as a state node.

\begin{figure}
    \centering
    \includegraphics[scale=0.8]{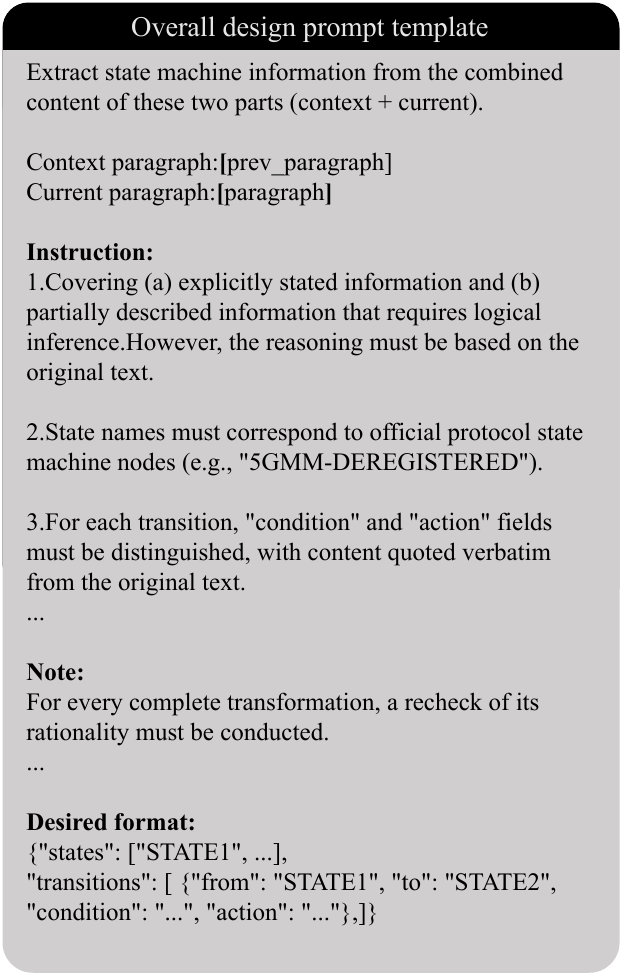}
    \caption{A unified basic prompt model for FSM}
    \label{unified basic}
\end{figure}

Unfortunately, the extraction of state tuples is often non-trivial. A major challenge arises from the presence of ambiguous or incomplete state names in the specifications themselves. For instance, in the NAS specification, the documents may describe states using vague terms such as ``any state" or ``some state," which lack precise definitions. Additionally, due to the absence of clear state matching, large language models may further generate so-called pseudo-states like ``Unknown" or ``Undefined" during the extraction process. Traditional extraction approaches—primarily based on keyword matching (e.g., detecting the word ``state" preceding a potential state name)—are insufficient to handle such ambiguities and the diverse expressions commonly found in protocol documents. To mitigate these issues, we incorporate explicit instructions into the prompt design to guide accurate state extraction, as illustrated in Figure~\ref{unified basic}. Specifically, the model is instructed to extract only standardized state names defined in the protocol, such as ``5GMM-REGISTERED," and to ignore non-standard or descriptive phrases that could introduce noise into the state set. Moreover, some states are presented in abbreviated forms, such as ``PLMN-SEARCH," whereas the full qualified state should be ``5GMM-REGISTERED.PLMN-SEARCH." To handle such cases, \toolname{} employs an associated search and multi-level joint reasoning strategy. During state extraction, it analyzes the semantic relationships among the target state, its surrounding context, and adjacent states to infer and reconstruct the complete state representation, thereby ensuring consistency and correctness.

\textbf{Transition Extraction.} Conditions and actions define the transitions in a state machine. Ideally, these elements are clearly stated in the specification. For example, a condition may be expressed as ``No 5GMM context has been established," with the corresponding action being ``The UE shall start the initial registration procedure." However, in practice, conditions and actions in 3GPP documents often appear within the same sentence, and their boundaries are not always well defined. This ambiguity often leads to confusion during extraction by large language models, mainly due to difficulties in understanding the logical relationship between the two components. To address this issue, as illustrated in Figure~\ref{unified basic}, we explicitly highlight the distinction between conditions and actions in our prompt design. A condition provides the premise for deciding whether a specific response should be triggered, while an action represents the behavior executed once the condition is satisfied. To further improve accuracy, we include few-shot examples \cite{parnami2022learning} to help the model learn and internalize this logical separation.

\textbf{Post-Processing.} In the prompt design stage, we incorporate specific restrictive instructions to guide the behavior of the large language models. However, due to the inherent limitations of these models, the generated outputs may occasionally deviate from the expected format or content. To address this, we introduce an auxiliary validation process to ensure output reliability. During this stage, we first perform structural validation by checking the correctness of the JSON format during parsing. Any outputs that fail to meet the required specification are automatically flagged and reported. In addition, we apply a series of post-processing measures, including removal of pseudo-states and empty states. These steps are implemented through predefined rules, enabling the system to effectively identify and eliminate undesired or noisy content.

\paragraph{\textbf{Prompt Optimization}}
SpecGPT aims to construct a holistic and comprehensive state machine representation. However, extracting the state machine based on individual text segments may lead to information loss, as certain transitions can span across segment boundaries. To address this limitation, we further develop prompt optimization strategies that leverage broader contextual information, enabling the model to infer and recover implicit state transitions.

\textbf{Collaborative Contextual Reasoning.} The large scale of 3GPP protocol specifications necessitates segmented processing, which makes the recognition window of large language models a critical factor influencing extraction performance. Although preprocessing merges content across segment boundaries, it does not fully achieve the construction of a coherent global state machine. To address this issue, SpecGPT incorporates a collaborative contextual reasoning mechanism. As illustrated in Figure~\ref{unified basic}, the prompt is explicitly designed to include relevant historical context alongside the current text segment, enabling the model to continuously capture protocol behavior across segments. This approach supports more accurate and consistent extraction of state machine elements by grounding the model’s reasoning in an extended context window.

\textbf{Integration of Explicit and Implicit Information.} Conventional structured parsing methods \cite{al2024hermes} primarily focus on the accurate extraction of explicitly stated information in protocol specifications, but often struggle to identify or capture deeper, latent logical relationships embedded within 3GPP texts. To address this limitation, we design prompt strategies that integrate explicit extraction with implicit reasoning, as illustrated in Figure~\ref{unified basic}. This approach not only enables the precise identification of clearly defined state machine elements, but also supports the inference of logically implied transitions that are not directly expressed in the text. To minimize the risk of hallucinations, our prompts enforce strict grounding in the original document content and require all inferred results to undergo a subsequent plausibility verification process.

\textbf{Cross-Reference Handling.} Cross-references are pervasive in 3GPP protocol specifications. If the content being referenced is not explicitly provided, it can result in contextual loss, as large language models are unable to access or retrieve the referenced material independently. To mitigate this issue, we design a simplified retrieval-augmented generation (RAG) \cite{gao2023retrieval} mechanism. Specifically, we allow multiple protocol texts to be input simultaneously and use the original document section numbers as prefixes to construct an integrated chapter-level mapping. This design supports retrieval-based input, ensuring that cross-referenced content is accurately located and incorporated into the prompt, thereby improving the completeness and reliability of extracted information. Compared to conventional RAG frameworks, our approach is lightweight, incurs minimal computational overhead, and requires no external knowledge base or embedding storage.

\subsection{Ensembling}
\label{Ensembling}
Despite the carefully crafted prompts informed by domain knowledge, the performance of large language models remains inherently constrained by limitations such as hallucination. To further enhance overall reliability and accuracy, we adopt a model ensemble strategy that aggregates outputs from multiple LLMs, thereby mitigating individual model biases and improving extraction robustness. Specifically, we select several mainstream LLMs from different providers, including GPT 4o and Gemini 2.5 Pro etc. For the same input task, we generate candidate outputs with identical prompts and standardized parameters (e.g., temperature settings). Due to architectural and behavioral differences among models, their interpretations and outputs may vary, necessitating manual alignment to remove obvious inconsistencies, such as omissions in state transitions. Additionally, since actions and conditions are extracted as original text spans from the source, different models may select different fragments. To address these discrepancies, we introduce a consensus mechanism based on the principle of state transition completeness, utilizing a correlation matching algorithm defined as follows.

Let \( T_i = (S_i^{\mathrm{init}}, A_i, C_i, S_i^{\mathrm{next}}) \) denote a state transition extracted by model \( i \), where \( S_i^{\mathrm{init}} \) and \( S_i^{\mathrm{next}} \) represent the initial and resulting states, \( A_i \) is the action text span, and \( C_i \) is the condition text span.

Two states \( S_i \) and \( S_j \) are considered matching if
\[
S_i = S_j.
\]

For actions and conditions, let \( \mathrm{span}(X) \) denote the text span of \( X \) measured in words. The overlap between two spans is defined as
\[
\mathrm{Overlap}(A_i, A_j) = \frac{|\mathrm{span}(A_i) \cap \mathrm{span}(A_j)|}{\min\big(|\mathrm{span}(A_i)|, |\mathrm{span}(A_j)|\big)} \geq \theta,
\]
where \( \theta \) is a predefined threshold indicating sufficient overlap. We set \( \theta \) as 0.75 in practice, which yielded the best performance in our evaluation. 

Similarly, for conditions,
\[
\mathrm{Overlap}(C_i, C_j) = \frac{|\mathrm{span}(C_i) \cap \mathrm{span}(C_j)|}{\min\big(|\mathrm{span}(C_i)|, |\mathrm{span}(C_j)|\big)} \geq \theta.
\]

Two transitions \( T_i \) and \( T_j \) are considered aligned if and only if all of the following hold:
\[
\begin{cases}
S_i^{\mathrm{init}} = S_j^{\mathrm{init}}, \\
S_i^{\mathrm{next}} = S_j^{\mathrm{next}}, \\
\mathrm{Overlap}(A_i, A_j) \geq \theta, \\
\mathrm{Overlap}(C_i, C_j) \geq \theta.
\end{cases}
\]

Based on the aforementioned algorithm, we align the outputs from different models and apply a majority voting strategy to integrate them, thereby obtaining the final finite state machine.

\section{Evaluation}
\label{evaluation}
We evaluate \toolname{} on different settings and address the following three research questions:
\begin{enumerate}[nolistsep, labelindent=\parindent,leftmargin=*, label=$\bullet$]
\item RQ1: How effective is our multi-model ensemble framework in extracting protocol semantics from real-world 3GPP documents?
\item RQ2: How does our approach compare to state-of-the-art protocol analysis tools?
\item RQ3: What is the practical cost and generalization ability of deploying our framework in real-world settings?
\end{enumerate}

\subsection{Settings}
We selected five advanced models from different vendors, namely GPT 4o \cite{chatgpt}, DeepSeek V3 \cite{deepseek}, Qwen Turbo \cite{qwen}, Claude Sonnet 4 \cite{claude}, and Gemini 2.5 Pro \cite{gemini}. These models represent the most advanced open-source and proprietary large language models available today \cite{lmarena}. The parameter sizes of these models range from tens to hundreds of billions. According to widely accepted recommendations for generating factual responses, we set the temperature parameter to 0.2 \cite{peeperkorn2024temperature}. All experiments were conducted on a machine equipped with 32GB RAM and Intel Core i7-14700 CPU.


\paragraph{\textbf{Groundtruth Construction}} Currently, there is still a lack of a comprehensive and authoritative ground truth dataset for 3GPP protocols. Existing datasets from prior work often suffer from limited protocol coverage and only capture explicit states or behaviors. To address this gap, we manually constructed a detailed ground truth state machine dataset for three widely used 5G core network protocols—NAS, NGAP, and PFCP. All protocols are based on Release 17 specifications. This effort involved over 210 person-hours of work by multiple domain experts, including cross-validation and iterative peer review to ensure protocol type coverage, state completeness, and strict compliance with 3GPP specifications. Our ground truth dataset significantly improves state transition coverage across mainstream protocols, providing a robust foundation for future research in protocol modeling, analysis, and verification.

\paragraph{\textbf{Evaluation Metrics}} We employ three widely used and objective measures: Precision, Recall, and F1-Score. Following standard definitions, precision is defined as $ \frac{TP}{TP+FP} $, Recall is given by $ \frac{TP}{TP+FN} $, the F1-Score is calculated as $ 2 \times \frac{\mathrm{Precision} \times \mathrm{Recall}}{\mathrm{Precision} + \mathrm{Recall}} $. Here, $TP$ denotes the number of true positives, $FP$ denotes false positives (i.e., the number of instances incorrectly predicted as positive), and $FN$ denotes false negatives (i.e., actual positive instances missed by the model). It should be emphasized that the F1-Score takes into account both $FP$ and $FN$, thereby serving as a crucial indicator for the overall performance of the model in this study.

\subsection{RQ1: End-to-End Performance}
We apply SpecGPT to extract state machines from the three protocols included in our ground truth dataset, namely the Release 17 versions of NAS, NGAP, and PFCP, in order to conduct a systematic evaluation of SpecGPT. Since the model outputs for conditions and actions are directly extracted from the source documents, their spans may not perfectly correspond to those in the ground truth. Therefore, we adopt an alignment strategy similar to that used in the ensembling process (Section \ref{Ensembling}). Specifically, a state machine transition is considered correct only if the states match exactly, and both the condition and action spans overlap beyond a specified threshold, which we set as 0.75 in experiments.



\paragraph{\textbf{Performance of State Extraction on NAS}}
The NAS ground truth we annotated includes a total of 18 distinct states and 179 transitions. All five models successfully and precisely extracted all these states within the NAS protocol, resulting in an F1-score of 100\%. This achievement highlights both the advanced understanding ability of large language models and the effectiveness of our specialized state extraction design.

\paragraph{\textbf{Performance of Transition Extraction on NAS}} Table \ref{Various models} presents the extraction performance of various large language models on state transitions within the NAS protocol. The row labeled "Ensemble" shows the aggregated results from the five models, reflecting \toolname{}’s overall performance in extracting state transitions. The effectiveness of transition extraction demonstrates \toolname{}’s comprehensive capability in the state machine extraction task. As indicated in the table, extraction performance varies among models despite using identical prompt configurations. Specifically, the models exhibit relatively low precision, ranging from 61.71\% to 80.39\%, while recall remains comparatively high, between 77.66\% and 92.55\%. This discrepancy is primarily due to the hallucination problem. Large language models tend to generate significantly more transition tuples than actually exist, leading to a higher false positive rate and consequently lower precision. However, this hallucination does not markedly affect the coverage of correct transitions, resulting in consistently high recall.
Furthermore, when ensemble decision strategies are applied across multiple models, the F1-score improves substantially, ranging from 5.85\% to 22.37\%. This improvement is attributed to the ensemble method’s ability to remove incorrect transitions caused by hallucinations in individual models, thereby boosting overall extraction performance.

Large language models were trained using 3GPP specification documents. We attempted to directly prompt these models to output state machines for different protocols, but the resulting F1-score was only 14.87\%. This outcome demonstrates the necessity of our carefully designed in-context prompts.


\begin{table}[htbp]
    \centering
    \caption{Comparison of the three indicators of various LLMs}
    \begin{tabular}{cccc}
         \hline
         Model&  Precision (\%)&  Recall (\%)&F1-score (\%) \\
         \hline
         Claude Sonnet 4&  80.39&  87.23& 83.67\\
         DeepSeek V3&  68.70&  84.04& 75.60\\
         Gemini 2.5 Pro&  70.00&  89.36& 78.50\\
         GPT 4o&  79.09&  92.55& 85.29\\
         Qwen Turbo&  61.71&  77.66& 68.77\\
         Ensemble&  91.86&  90.43& 91.14\\
         \hline
    \end{tabular}
    \label{Various models}
\end{table}

\paragraph{\textbf{Performance of Transition Extraction on PFCP and NGAP}}





\begin{table}[htbp]
    \centering
    \caption{Indicator data at different levels of different protocols}
    \begin{tabular}{cccc}
         \hline
         Protocol&  Precision (\%)&  Recall (\%)&F1-score (\%) \\
         \hline
         PFCP-all&  85.71&  90.00& 87.80\\
         PFCP-asscoiation&  85.71&  84.04& 80.00\\
         PFCP-session&  92.30&  92.30& 92.30\\
         NGAP-all&  70.71&  67.96& 69.31\\
         NGAP-NCG&  91.67&  71.43& 80.29\\
         NGAP-PSR&  72.92&  68.63& 70.71\\
         NGAP-UCM&  57.43&  64.89& 60.93\\
         \hline
    \end{tabular}
    \label{protocols}
\end{table}
Table \ref{protocols} summarizes the performance of SpecGPT in extracting state transitions within the PFCP and NGAP protocols. To enable a more detailed evaluation of SpecGPT’s capabilities, both protocols were further subdivided for fine-grained assessment across distinct protocol components. Specifically, for the PFCP protocol, we separated the association establishment and maintenance functions (association layer) from session management and forwarding rule control (session layer), treating them as two independent layers.
For the NGAP protocol, the subdivision was based on functional granularity and control scope, resulting in three hierarchical layers from the network to the service bearer level: the Next-Generation Core (NGC) Connection Global layer, describing the global state of the NGC interface; the UE Context Main (UCM) layer, responsible for signaling and state management at the UE level; and the PDU (Protocol Data Unit) Session Resource layer, focusing on PDU session resources for each UE.
Experimental results demonstrate that SpecGPT achieves excellent extraction performance on the PFCP protocol both overall and within each individual layer, with particularly precise recognition in the association and session layers. In contrast, the NGAP protocol is more complex and exhibits certain ambiguities in its specification, increasing the difficulty of automatic extraction. Nonetheless, its overall extraction performance still satisfies the basic requirements for practical application. 

\subsection{RQ2: Comparison with the SOTA}
To evaluate the effectiveness of SpecGPT, we further conduct a comparative evaluation against Hermes \cite{al2024hermes}, the SOTA tool designed to automatically extract protocol finite state machines (FSMs) from 3GPP specification documents.

Since the Hermes paper only reports experimental results on the 5G NAS and RRC protocols, we restrict our comparison with Hermes to the 5G NAS protocol. The main limitation of Hermes lies in its explicit extraction, which is restricted to fully articulated transitions explicitly stated in protocol specification texts, and does not sufficiently account for the semantic logic and implicit information inherent in the natural language formation of specifications. In contrast, SpecGPT is capable of capturing implicit state transitions embedded in the protocol specifications. Hermes does not provide the actual state machines it constructed. According to the evaluation results reported in its paper, Hermes achieves an accuracy of 81.39\% for actions and 86.40\% for conditions on the 5G NAS protocol. In comparison, SpecGPT achieves higher accuracies of 86.41\% and 92.94\%, respectively. It is worth noting that the ground truth used by SpecGPT is more comprehensive in terms of both the number of states and transitions. Therefore, SpecGPT demonstrates superior overall performance on the 5G NAS protocol compared to Hermes.

\begin{table}[htbp]
    \centering
    \caption{The effects of different tools in labeling tags}
    \begin{tabular}{cccc}
         \hline
         Tools&  Precision (\%)&  Recall (\%)&F1-score (\%)\\
         \hline
         NEUTREX-Labeled&  64.30&  66.13& 65.20\\
         NEUTREX-Unlabeled&  66.88&  68.79& 67.82\\
         LLM&  87.46&  90.40& 88.90\\
         \hline
    \end{tabular}
    \label{different tools}
\end{table}

In addition, Hermes introduces a dedicated neural constituency parser, NEUTREX, to annotate transition components such as states, conditions, and actions in 3GPP specification documents. To assess the effectiveness of large language models (LLMs) on the same task, we use the ground truth provided by Hermes, which includes human-annotated transition components extracted from the specifications. Using this dataset, we apply LLMs to perform the annotation task without any task-specific training.
The results are shown in Table \ref{different tools}. In this table, NEUTREX-Unlabeled denotes NEUTREX’s performance on the detected spans without any consideration of the labels assigned to them, while NEUTREX-Labeled take into consideration the labels assigned.
Overall, LLMs outperform NEUTREX on this task. These findings underscore the advantages of our end-to-end approach, which achieves higher accuracy without relying on a dedicated parser or training process. Figure 4 presents a breakdown of the performance of different large language models on each type of transition component in this task. All five models demonstrate strong performance across the annotation of individual components. This also highlights the potential of LLMs to simplify and enhance automated protocol analysis by eliminating the need for complex intermediate models.

\begin{figure}[htbp]
    \centering
    \includegraphics[width=1\linewidth]{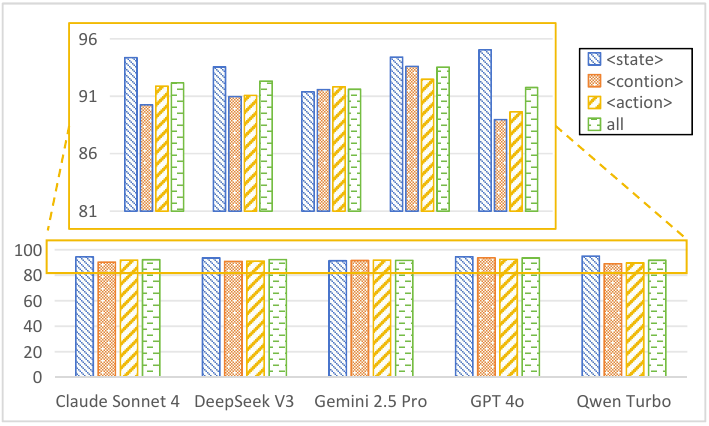}
    \caption{The effectiveness of each model in labeling the tags}
    \label{effectiveness of each model}
\end{figure}

\subsection{RQ3: Cost and Generalization Ability}

\paragraph{\textbf{Cost}} We assess the practical cost of our approach from two perspectives: monetary expense and runtime efficiency. Specifically, we measure the number of tokens consumed and the processing time required by SpecGPT when invoking large language models and executing subsequent processing pipelines. Regarding monetary cost, token usage is the main factor, which largely depends on the length of the 3GPP specification documents. Table \ref{costs} summarizes the detailed token consumption for each model, showing comparable usage across the evaluated models. Based on the respective API pricing, we estimate the cost per run to be approximately \$2.7 for NAS, \$1.6 for NGAP, and \$1.5 for PFCP. In terms of runtime efficiency, the primary influence is the type of LLM employed. Through optimized input window scheduling and segmentation strategies, we have significantly reduced the processing time for a single document from several hours to mere minutes. Additionally, preprocessing and comparative testing typically require less than one minute. Detailed performance metrics are also presented in Table \ref{costs}. Considering that protocol state machine extraction is usually performed only once per document, the overall time cost is acceptable in practical scenarios. With ongoing advancements in foundation models, we expect both monetary costs and processing time to further decrease.


\begin{table}[htbp]
    \centering
    \caption{The costs and efficiencies of each model under each protocol}
    \begin{tabularx}{\linewidth}{ccccc} 
        \hline
        Protocol & Model& \makecell[c]{\# Input \\ Tokens} & \makecell[c]{\# Output \\ Tokens} & Time (minute) \\
        \hline
        \multirow{5}{*}{NAS} 
            & Claude Sonnet 4  & 221,829 & 50,347 & 32 \\
            & DeepSeek V3      & 233,517 & 75,239 & 117 \\
            & Gemini 2.5 Pro   & 236,751 & 74,985 & 202 \\
            & GPT 4o           & 235,984 & 94,202 & 20 \\
            & Qwen Turbo      & 234,894 & 97,863 & 57 \\ \hline
        \multirow{5}{*}{NGAP} 
            & Claude Sonnet 4  & 130,490 & 29,358 & 20 \\
            & DeepSeek V3      & 136,578 & 43,857 & 69 \\
            & Gemini 2.5 Pro  & 136,875 & 43,221 & 119 \\
            & GPT 4o          & 136,560 & 55,461 & 12 \\
            & Qwen Turbo      & 136,423 & 57,400 & 35 \\ \hline
        \multirow{5}{*}{PFCP} 
            & Claude Sonnet 4  & 118,904 & 26,721 & 18 \\
            & DeepSeek V3      & 124,551 & 39,826 & 61 \\
            & Gemini 2.5 Pro   & 124,686 & 39,801 & 113 \\
            & GPT 4o         & 124,558 & 49,577 & 10 \\
            & Qwen Turbo       & 124,496 & 51,996 & 31 \\ \hline
    \end{tabularx}
    \label{costs}
\end{table}
\paragraph{\textbf{Generalization Ability}} We further assess the generalization ability of SpecGPT on a different version of the specifications, as real-world applications typically require re-extracting state machine models following specification updates. Specifically, the evaluation is performed using the Release 15 (R15) version of the NAS protocol specification. SpecGPT successfully extracted 142 state transitions from the Release 15 (R15) version of the specification, representing a reduction of approximately 20\% compared to Release 17 (R17). Notably, the total number of states remained largely unchanged. This outcome is consistent with expectations, as R15 serves as the initial 3GPP standard defining 5G NR and 5GC, whereas R17 introduces subsequent enhancements, including new features, commercial deployment scenarios, and expanded network capabilities, which collectively contribute to increased complexity in the protocol’s state machine. The results of the experiment validate the generalization ability of our tool, indicating its effectiveness in handling multiple versions of the protocol specifications.



\section{Discussion}
\label{discussion}

\subsection{Limitations and Future Work}
Although SpecGPT achieved an F1-score of 91.14\% in extracting protocol state machines from 3GPP documents and significantly outperformed the existing SOTA tool, it still suffers from false positives and false negatives due to inherent limitations such as hallucinations in large language models. This challenge is common across many LLM-based approaches \cite{huang2025survey, xu2024hallucination}. Addressing this issue requires progress in two key directions. The first is the continual improvement of foundation model capabilities. The second involves the integration of emerging techniques such as retrieval-augmented generation \cite{gao2023retrieval} and self-consistency \cite{wang2022self} to reduce hallucinations and improve the reliability of model outputs. As large language model technologies continue to evolve rapidly, our extensible multi-model ensemble framework is well positioned to benefit from these advancements. This will enable continuous improvement in state machine extraction and ultimately support the goal of generating complete and accurate protocol state machines. In addition, this work focuses exclusively on protocol state machines. However, protocol specification documents contain a wealth of other valuable information, such as protocol fields and potential security hazards, which play a critical role in various security-related tasks. We leave the extraction and utilization of such information to future work.

\subsection{Utility of \toolname{}}
This work focuses on the automatic extraction of protocol state machines from 3GPP specifications. The proposed framework can facilitate a variety of downstream tasks, especially in protocol verification and testing, and the enrichment of technical specifications.

\paragraph{\textbf{Protocol verification and testing}} Most existing tools for protocol verification and testing rely on hand-crafted state machines, which are time-consuming to build, error-prone, and difficult to maintain. With frequent updates to protocol specifications, manually constructed models often become outdated quickly. SpecGPT can automatically extract state machine models from the specifications, significantly reducing the manual effort required and improving the maintainability and accuracy of existing protocol analysis research.

\paragraph{\textbf{Enrichment of 3GPP specifications}}  The main reason researchers have to manually construct state machine models is that the 3GPP, as the maintainer of protocol specifications, provides the standards only in natural language, rather than in more structured forms such as state machines, which are better suited for a wide range of downstream tasks. This choice is influenced by various factors, but a major constraint is the substantial manual effort required to create and maintain such formal representations. Our tool addresses this gap by automatically extracting protocol state machines from the specifications, significantly reducing human workload and enabling structured, machine-readable representations that can greatly support tasks such as verification, testing, and protocol evolution.

\section{Related Work}
\label{related}
\subsection{Cellular Security}
The complexity and continuous evolution of cellular network protocols have driven growing interest in automated security analysis and verification approaches \cite{akon2025control, basin2018formal, hussain20195greasoner, hussain2018lteinspector, karim2021prochecker, dongcorecrisis, yang2024feedback, bennett2024ransacked, kim2019touching,shaik2017practical}. Karim et al. \cite{karim2021prochecker} introduced ProChecker, a framework that combines dynamic testing with static instrumentation to detect logical flaws in 4G LTE protocol implementations.
Akon et al.\cite{akon2025control} proposed CoreScan, a decomposition-based framework with assume-guarantee reasoning to verify 5G access control, uncovering five new privilege escalation vulnerabilities. Dong et al.\cite{dongcorecrisis} proposed CoreCrisis, a stateful fuzzing framework that learns an FSM to guide state-aware mutation for testing 5G core implementations. It discovered 7 specification deviations and 13 critical vulnerabilities. While existing works have advanced automated security analysis of cellular protocols, automation in state machine modeling remains limited—especially for the complex, dynamic interfaces in 5G and beyond.

\subsection{Machine Learning Based Document Analysis}
Machine learning, particularly NLP, is increasingly used to extract structured protocol specifications from documents and code, supporting automated protocol parsing and state modeling \cite{chen2021bookworm, chen2022forest, pacheco2022automated, al2024hermes,rashid2024state}, \cite{CellularLint2024, Wang2023}. Chen et al.\cite{chen2022forest} introduced CREEK, which uses a fine-tuned BERT model and Positive-Unlabeled learning to classify security-relevant 3GPP Change Requests, and successfully uncovered multiple exploitable security vulnerabilities. Ishtiaq et al.\cite{al2024hermes} proposed Hermes, which combines a neural parser, a domain-specific language for intermediate representation, and logic-driven FSM generation to automatically extract state machines from 3GPP specifications using NLP. Rahman et al.\cite{CellularLint2024} applied NLP to identify design-level inconsistencies in specification documents and assessed their potential impact on implementation choices. However, prior methods rely on annotated data and domain-specific training. Whether the capabilities of large language models can be leveraged for efficient protocol recognition is the focus of this work.

\section{Conclusion}
\label{conclusion}

In this work, we present \toolname{}, a novel framework for automatically extracting protocol state machines from 3GPP specifications using large language models. By combining domain-informed prompt engineering, chain-of-thought reasoning, and ensemble techniques, \toolname{} effectively addresses the complexity, scale, and ambiguity inherent in cellular protocol documents. Our evaluation on NAS, NGAP, and PFCP protocols demonstrates that \toolname{} achieves higher accuracy compared to prior approaches, highlighting the potential of LLMs in automating protocol analysis. The state machines extracted by our framework serve as structured, interpretable representations of protocol behavior, which can further facilitate downstream tasks such as automated testing and formal verification, ultimately enhancing the robustness and security of modern communication protocols.

\bibliographystyle{IEEEtran}
\bibliography{reference}

\end{document}